# Structured Unit Testable Templated Code for Efficient Code Review Process

Amol S Patwardhan

Department of Mechanical and Industrial Engineering, LSU, apatwa3@lsu.edu

## Abstract

Background: Modern software development teams are distributed across onsite and off-shore locations. Each team has developers with varying experience levels and English communication skills. In such a diverse development environment it is important to maintain the software quality, coding standards, timely delivery of features and bug fixes. It is also important to reduce testing effort, minimize side effects such as change in functionality, user experience or application performance. Code reviews are intended to control code quality. Unfortunately, many projects lack enforcement of processes and standards because of approaching deadlines, live production issues and lack of resource availability.

Objective: This study examines a novel structured, unit testable templated code method to enforce code review standards with an intent to reduce coding effort, minimize revisions and eliminate functional and performance side effects on the system. The proposed method would also result in unit-testable code that can also be easily rolled back and increase team productivity.

Method: The baseline for traditional code review processes using metrics such as code review duration, bug regression rate, revision count was measured. These metrics were then compared with results from the proposed code review process that used structured unit testable templated code. The performance on 2 large enterprise level applications spanning over 2 years and 9 feature and maintenance release cycles was evaluated.

Results: The structured unit testable templated code method resulted in a decrease in total code review time, revision count and coding effort. It also decreased the number of live production issues caused by code churn or side effects of bug fix when compared to traditional code review process.

Conclusion: The study confirmed that the structured unit testable templated code results in improved code review efficiency. It also increased code quality and provided a robust tool to enforce coding standards in a cross-continent software maintenance team environment. It also relieved core resources from code review effort so that they could concentrate more on newer feature development.

## 1. Introduction

Code review is a vital step in the software development process because it ensures code quality at an early stage of release lifecycle and also provides an opportunity to inculcate coding best practices. Code review as a quality control tool has been identified in the 1980s (Ackerman, Fowler & Ebenau, 1984) and (Ackerman, Buchwald & Lewski, 1989). Fagan, 1976 suggested a formal process involving group reviews, meetings for code reviews. Votta, 1993 investigated whether a formal, time consuming and meeting oriented code review is needed. Today software development teams are spread across several onshore (within same country) and offshore (international) locations. The teams often consist of wide range of programming and communication skills in terms of experience and culture. In such a diverse setting accountability and ensuring code quality becomes very important. The geographical distribution also makes it harder to implement the formal code review process outlined by Fagan. Researchers (Bacchelli & Bird, 2013) have identified the challenges to the code review process and the expectations from a successful code review. The researchers reported that many organizations struggle to execute and enforce the code review process primarily because of following reasons:1) Low reviewer participation, 2) Poor knowledge of code context, 3) Pressure to meet deadlines. According to the study the main requirements and expectations of developers and management from efficient and successful code review process are:1) Short execution time, 2) Maintain coding standards, 3) Minimize performance impact, 4) Minimize breaking change, 5) Minimize functional side effects, 6) Optimal usage of reviewer time, 7) Inculcate good coding habits, 8) Knowledge transfer.

The software development process is shifting more and more towards agile development and continuous deployment. There is an increasing need for shorter development cycles and quicker code reviews. Moreover, newer flexible architectures and technologies such as real time embedded (Patwardhan, 2006), xml entities based (Patwardhan & Knapp, 2014), (Patwardhan, 2016), self-contained plugins (Patwardhan & Vartak, 2016) and Kinect based systems (Patwardhan & Knapp, 2013) are constantly being adopted and implemented. Such systems require in depth knowledge about the system for context aware and rapid code reviews and traditional formal methods can cause delays. Beller et. al, 2014 have examined the modern code review process in open source software projects. The modern code review process has the following characteristics: 1) An informal review process, 2) Increase use of code review tools, 3) Popular among well-known companies. (Laitenberger, 2002), (Johnson, 2006), (Porter, 1996) have examined various code review methods, collaboration processes and software inspection workflows which are meeting based and do not suit well for the modern agile software development. The researchers examined effects of team size, number of reviewers, sessions on code quality.

Researchers have developed tools like groupware and scrutiny to improve the code review participation and reduce code review time (Brothers, Sembugamoorthy & Muller, 1990), (Gintell et. al, 1993). Baysal et. al, 2013 have shown that the organizational and personal factors have an influence on the completion time of code review process. Various metrics and factors influencing code review such as code coverage, reviewer participation and expertise have been examined by researchers (Kemerer & Paulk, 2009), (Mantyla & Lassenius, 2009), (McIntosh et. al, 2014), (Sutherland & Venolia, 2009). An extensive research on open source projects and the code review process has been done by researchers (Rigby, German & Storey, 2014), (Rigby et. al 2012).

As a result, the primary contributions of this paper are:

1) Develop a novel coding instrument called structured, unit testable templated code.
2) Provide empirical evidence that the proposed method improves code review efficiency and can be used to augment modern code review process.
3) Improve reviewer participation.

Maintenance releases are routine software improvements containing fix for list of high priority bugs. The frequency varies from organization to organization, depending on complexity of the software and the business domain. Hot fixes are used to release extremely urgent and critical issues reported by users in live production environment. Such bug fixes are commonly called code or data patches and are included either in a maintenance releases or as a hot fix. For this research a structured unit testable templated code for the data layer was created. This approach enabled focusing on a specific problem area to test the hypothesis that a structured unit testable templated code can improve the efficiency of the code review process. The same principle can be easily extended to the business or User interface layer. Even though there are many different scripting languages (c#, java, vb, php, python, JavaScript) and structured query languages (MSSQL, MySQL, oracle), the underlying programming constructs (variable declaration, loops, conditional statements, transactions, error handling, object oriented programming concepts) essentially remain the same and as a result the templates can be made available in any programming language.

Writing code in the relational database layer using SQL requires a different mindset as compared to writing object oriented code. In the SQL world the programmer has to think in terms of sets and should know how to effectively use joins, indexes and prudently fetch and manipulate data. In contrast writing code in the business and the user interface layer requires application of object oriented programming language principles and involves heavy usage of loops, object instances and control flow. Programmers used to object oriented scripting languages, when assigned with writing SQL scripts struggle to adapt while dealing with data sets and relational data optimization strategies. They have limited understanding about SQL constructs (indexes, transactions, merge, joins) and tend to make a lot of mistakes. This results in higher number of code review iterations and inability to identify system wide implication of the code. Additionally, a lot of poorly written code also stems from low code reviewer participation and engagement from senior programmers. Programmers are either busy with on-going feature development and seldom engage in detailed, intellectual discussion about the code or tend to focus on formatting issues and syntactical errors that are obvious and easy to find.

## 2. Experimental Design

The baseline for traditional code review processes using metrics such as code review duration, bug regression rate, revision count was measured. These metrics were then compared with results from the proposed code review process that used structured unit testable templated code. The performance on 2 large enterprise level applications spanning over 2 years and 9 feature and maintenance release cycles was evaluated.

The first enterprise level application (internal code P1) was a web application built using ASP.NET C#, .NET Framework 4.0 and used MS SQL as the database. The architecture adopted for the product was 3-Tier architecture, implemented using web forms (presentation layer), business controllers in the processing layer and data access controllers in the data layer. The software development team was based in south east region of United States (2 Architects, 3 senior developers and 4 junior-mid level developers) and the support teams were located in offshore locations such as Mexico (1 team lead, 2 senior and 3 junior developers), East Europe (1 team lead, 3 senior and 5 junior developers) and Chile (1 team lead and 3 developers). The code was maintained using team foundation system (TFS 2010).

The second enterprise level application was a web application built on micro-services architecture. The application was developed using ASP.NET C#, .NET Framework, MVC, WCF services and jQuery on the client side. The software development team was based in west coast United States (1 Architects, 5 team leads, 11 mid-junior level developers) and offshore support team in Mexico (1 team lead, 5 senior and 2 junior developers) and offshore support team in India (1 team lead, 5 developers). The code was maintained using TFS 2010. Thus the experiments were conducted on projects with two completely different architectures (P1 used 3-Tier and P2 used micro-services).

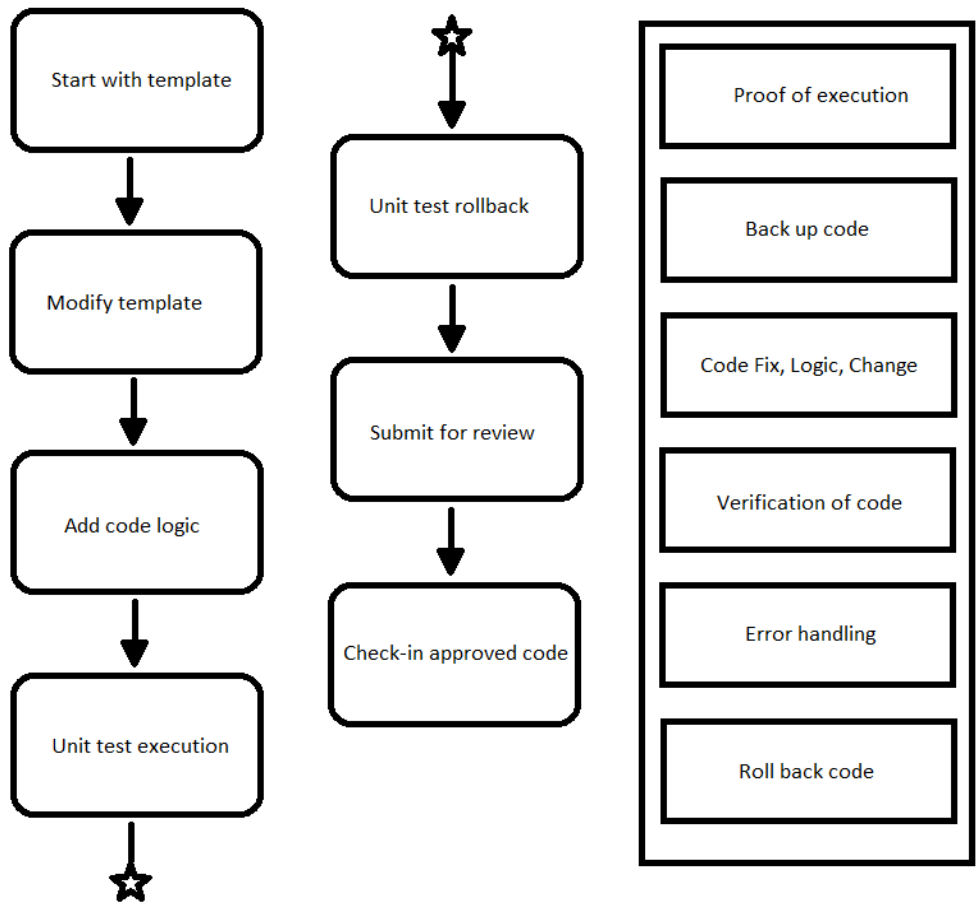

Fig. 1. Code review process workflow and structured unit testable code template block diagram.

The metric for the releases prior to using template were taken from two releases of each project and will be referred as pre-template releases. The internal pre-template release codes for project P1 were 5.4 and 5.6 and the internal pre-template release codes for project P2 were 5.8 and 6. The readings obtained for the pre-template releases established the baseline for the experiment. The development teams were provided a structured unit testable templates and a guide explaining the process. After the teams had understood and adopted the templates in the code review process for a total of 5 releases across two projects, the metrics were obtained again and compared with the baseline readings. Querying for the various metrics (revision history, comments counts, duration of review) was done using TFS. The releases that used the structured unit testable templated code were internally named release 7.13, 7.14 for project P1 and 7.15, 7.16, 7.17 for project P2.

For the purpose of this study SQL templates were used. The structure of the template is provided in the reference material. Figure 1 shows the code review process followed by the participating development teams and the structure of code template.

## 3. Results

Revisions is the iterations between reviewer and coder to make code corrections based on feedback. Lower number of revisions indicates quicker turn-around time and increased diligence from the programmer prior to submission of code for review. The number of revisions needed to ensure code-quality and adherence to standard was measure for the pre and post template releases.

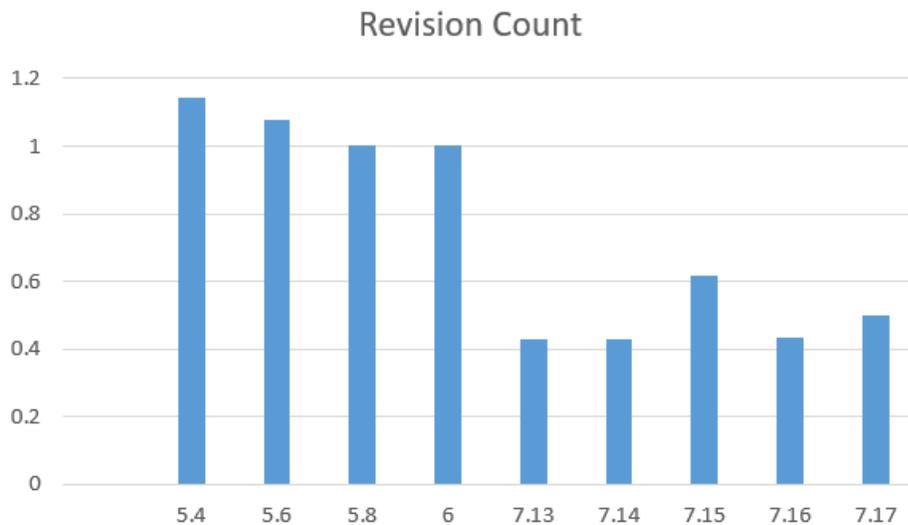

Fig. 2. Average revision count per release

Compared to the pre-template releases (5.4-6), the average revision count decreased for the post-template releases (7.13-7.17). The number of revisions needed prior to the templates was at least 1 or above, whereas post-template revision count was less than 1 (since the submitted code was correct and required no revisions because of conformance to the template).

The number of comments during a code review was measured for the pre and post template releases. The average comments for pre-templates was above 3 comments per requested code review. The average comments for post-template was below 3 comments per requested code reviews.

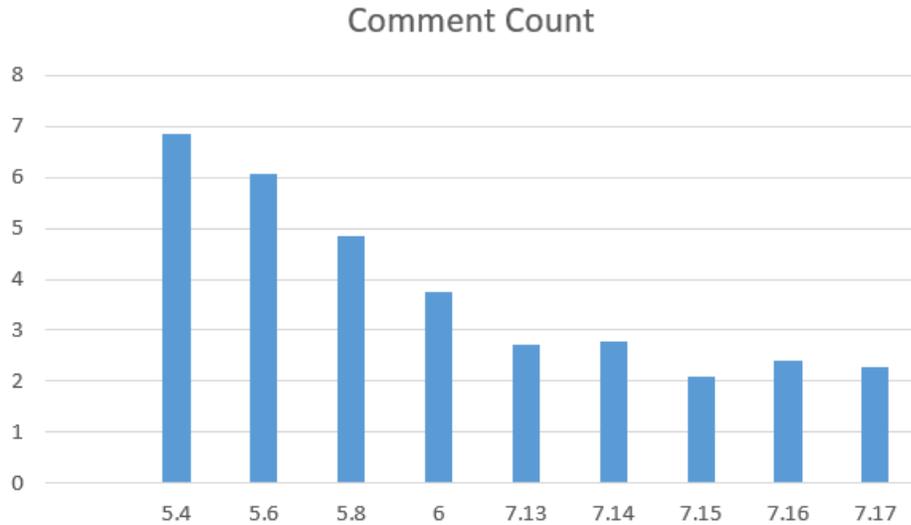

Fig. 3. Average code review comments per release

The decrease did not mean reduced reviewer participation and was actually an improvement in process efficiency. The templates resulted in higher adherence to coding standards and reduction in revisions. This resulted in decrease in comments per code review.

The number of bugs (software defects) regressed in pre-template releases was compared to the post-template defects in the quality assurance (QA) environment.

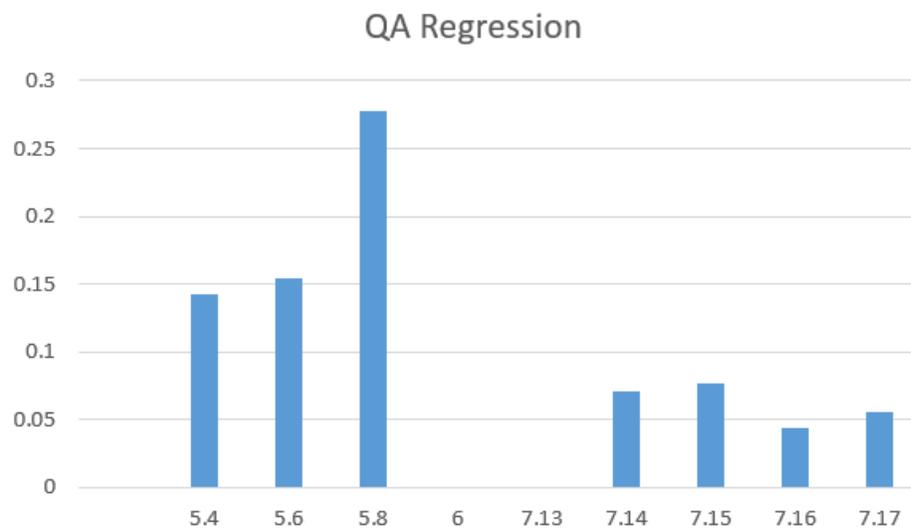

Fig. 4. Average bugs regressed per release in QA.

The QA bug regression count was higher than 0.1 for the pre-template release and lower than 0.1 for the post-template release. This indicated an improvement in code quality.

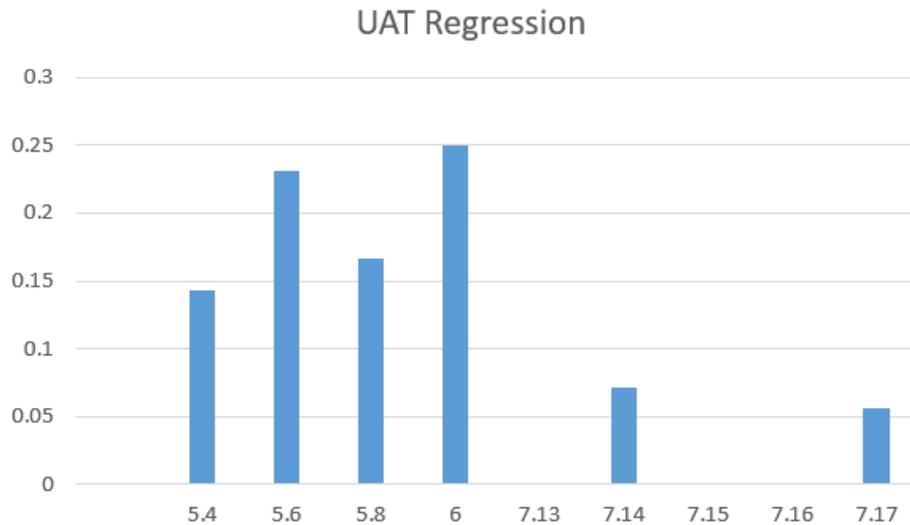

Fig. 5. Average bugs regressed per release in UAT.

The number of bugs (software defects) regressed in pre-template releases was compared to the post-template defects in the user acceptance testing (UAT) environment. The UAT bug regression count was above 0.1 for the pre-template release and below 0.1 for the post-template release. This indicated an improvement in code quality and reliability on code sign off from the QA environment.

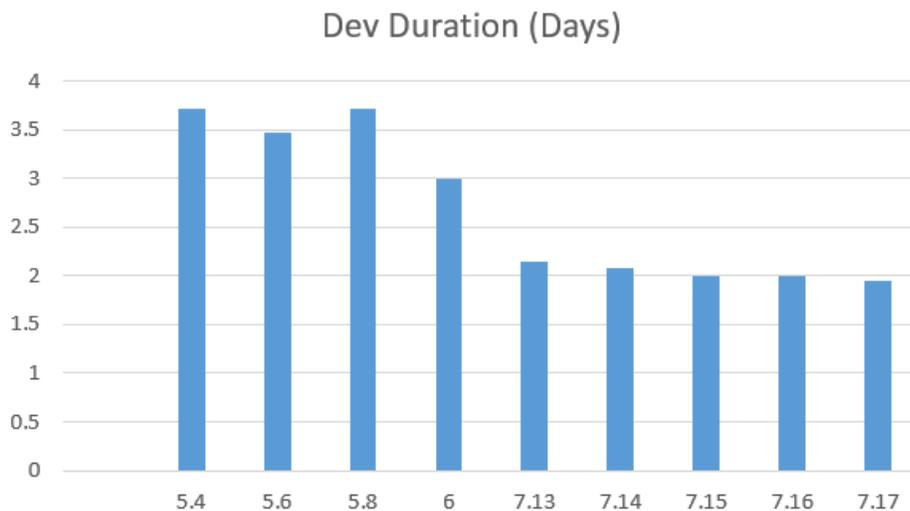

Fig. 6. Average development duration time in days per release.

Development time for fixing a bug decreased below 2.5 days for the post-template releases. This indicated that the structured approach towards code development improved coding efficiency.

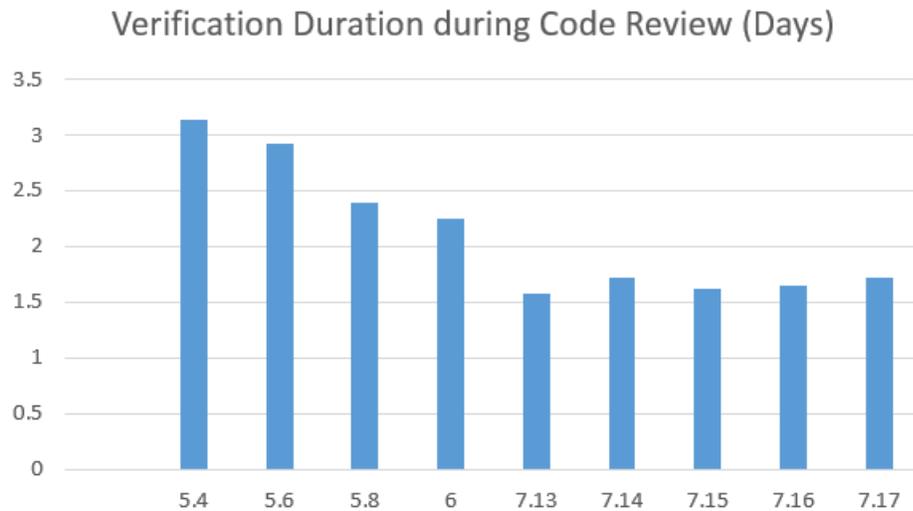

Fig. 7. Verification duration in days per release.

The code review process involves verification whether the fix executes properly and solves the problem. The average verification duration for the pre-template releases was above 2 days and decreased below 2 days for the post-template releases. This indicated an improvement in verification process. The structured unit testable code allowed the reviewers to look at the execution results within the submitted files and verify quickly without having to spend time on recreating the issue or setting up the pre-conditions for the issue.

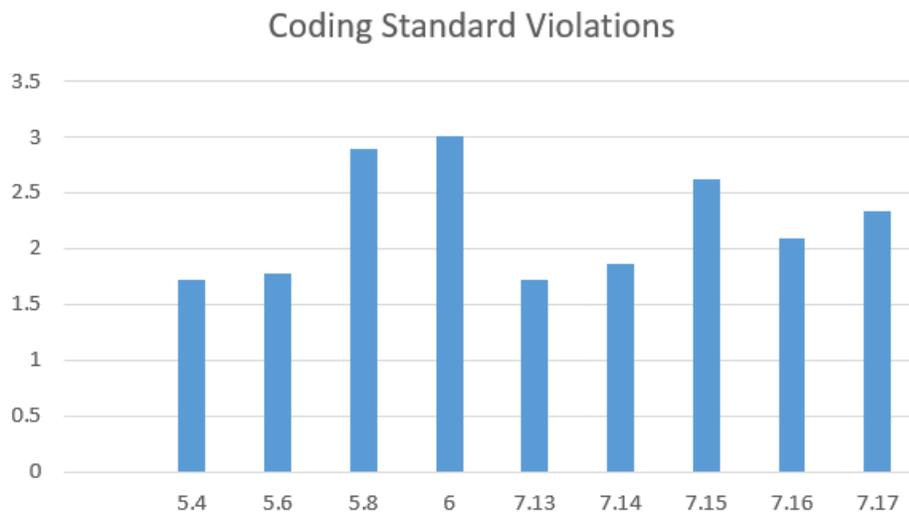

Fig. 8. Average coding standard violations per release

The coding standard violations count stayed above an average 1.5 per release for pre and post-template releases. An explanation for this could be that the improvement in the code review process because of the templates allowed the reviewers to focus more on the code quality resulting in a higher coding standard violation count. The pre-template releases had 2 releases with a high level of coding standard violations but the overall coding standard violation count did not fall below a specific threshold because of the new structured unit testable templates.

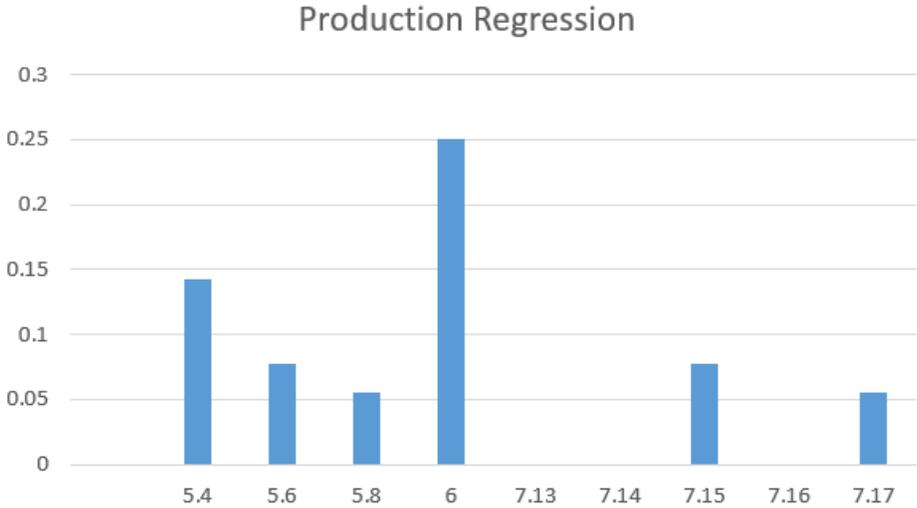

Fig. 9. Average production bug regression count per release

The number of fix reported to have failed in production decreased in the post-template releases with releases 7.13, 7.14 and 7.16 reporting 0 failures in productions for the fixed bugs. This indicated high robustness of the released software quality and an improvement in code review process.

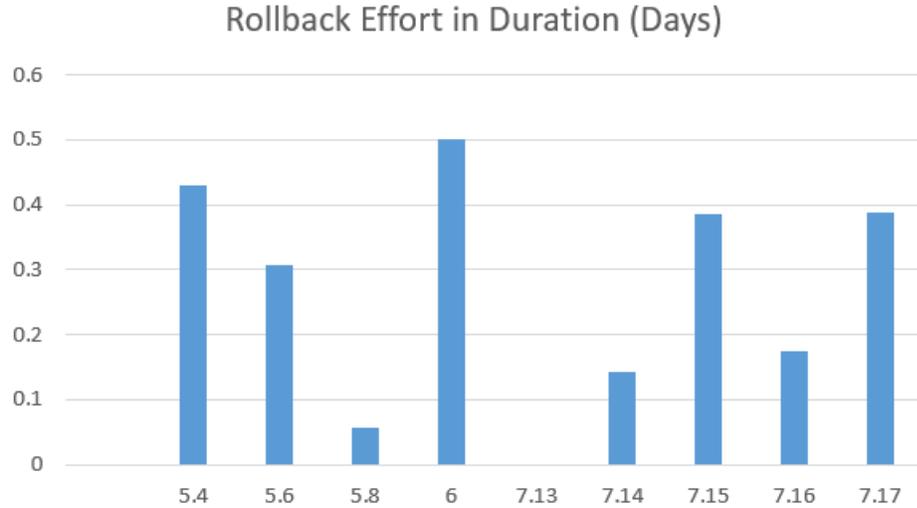

Fig. 10. Average rollback effort in days per release

Sometimes a feature or a fix for a software defect has to be rolled back for reasons such as failed test, missed delivery deadline and removal from the release, changes in regulatory requirements and other unanticipated reasons. The code fix released should support roll backs with minimum steps and complexity. The above metrics are across all three (QA, UAT and Production) environments. The number of days taken for rolling back a fix did not show a clear threshold difference between pre and post template release. This can be explained by the fact that although the templates provided a structured way to roll back the changes there were challenges in terms of coordination with deployment team, production support team and lack of understanding about the data and code among the non-development resources.

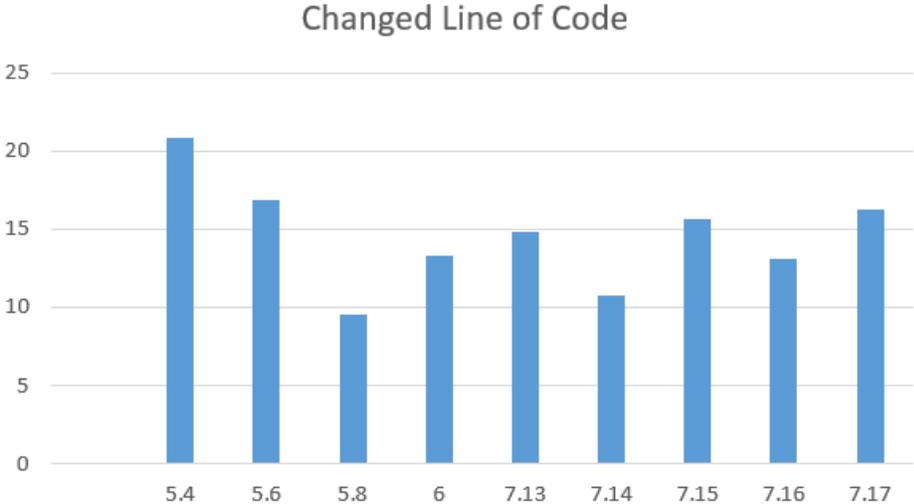

Fig. 11. Average lines of code per release.

The pre-template releases showed average lines of code ranging from 21 to 9 which indicated an un-restrained coding style. On the contrary the average lines of code for post-template was above 10 lines but did not go above 16. This showed that the structured templates enforced a consistent coding style.

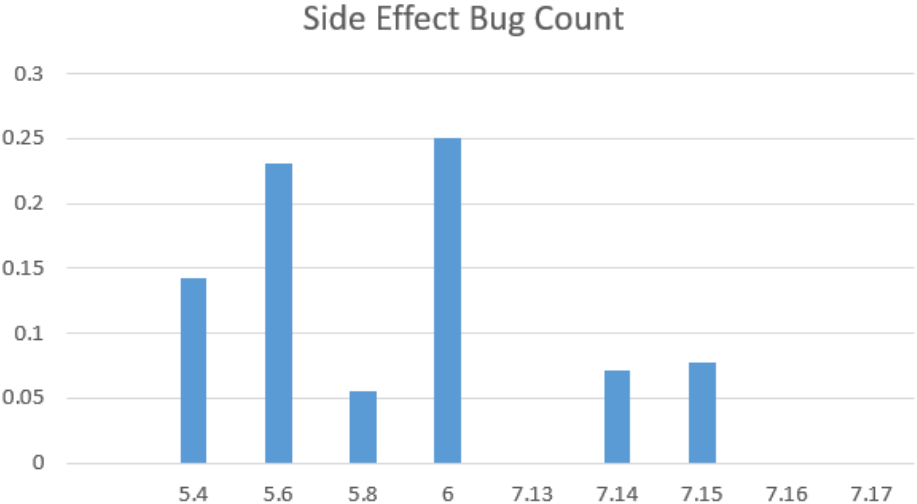

Fig. 12. Side effect bug count per release.

Sometimes a code change causes undesired side effects in user experience, functionality or regression of previous bug fix and features. The quality assurance team searched in existing list of bugs and associated (linked items in TFS) such issues with the current bug being tested. The average number of side effect bugs for the pre-template releases decreased during the post-template release below 0.75. This indicated an improvement in code quality and code review process.

## 4. Conclusion

The structured unit testable templated code provided a guided approach towards a reliable and efficient code review process. The structured templated code gave the programmers clarity in terms of the layout of their code and instead focus on the logic for bug fix and feature development. The process allowed the reviewers to focus on the context, actual issue, code logic without having to spend too much time in unit testing, verifying the fix, recreating the issue or setting up the environment.

Overall the templates improved the code quality and code review process efficiency. It also proved to be an effective tool to enforce code review process and standards across teams located in different continents and having varying level of coding skills and English language speaking skills.

As a future scope the templates need to be implemented and tested against a wider variety of programming languages and organizations of various size and maturity. It would be interesting to see how the templates work in a start-up or a development shop implementing a new software product as opposed to mature software products (projects P1 and P2 were well into their $7^{th}$ and $10^{th}$ year development cycle) that are mostly in the maintenance phase.